\documentstyle[prl,twocolumn,aps,times,epsf]{revtex}

\newcommand{\degree}{\mbox{$^\circ$}}

\newcommand{\micron}{\mbox{$\,\mu$m}}

\begin{document}

\twocolumn[\hsize\textwidth\columnwidth\hsize\csname @twocolumnfalse\endcsname

\title{A Self-Assembled Microlensing Rotational Probe}
\author{James P. Brody and Stephen R. Quake}


\address{Department of Applied Physics\\California Institute of Technology,  MS 128-95\\ Pasadena, CA 91125}
\date{\today}
\maketitle

\begin{abstract}
  A technique to measure microscopic rotational motion is presented.
  When a small fluorescent polystyrene microsphere is attached to a
  larger polystyrene microsphere, the larger sphere acts as a lens for
  the smaller microsphere and provides an optical signal that is a
  strong function of the azimuthal angle.  We demonstrate the
  technique by measuring the rotational diffusion constant of the
  microsphere in solutions of varying viscosity and discuss the
  feasibility of using this probe to measure rotational motion of
  biological systems.
\end{abstract}

\pacs{28.5,32.2,453.2}

\vskip1pc]





The commercial development of precision microspheres with well
controlled sizes and various protein coatings has enabled a number of
new devices and experimental techniques, from being the component
parts in the fabrication of photonic crystals~\cite{miguez98} to their
use as ``handles'' for optical tweezers in the study of single
protein~\cite{svoboda93,finer94} and DNA molecules~\cite{quake94}.  Here
we describe the use of microspheres with streptavidin-biotin coating
to self assemble a microscopic lensing system that can in turn be used
to measure rotation.

Previous biological experiments to measure rotational motions used a
``tethered cell'' assay in which a single flagellum is attached to a
surface and the cell rotates~\cite{berry97,berg74,silverman74}, or a
``rotating filament'' assay by attaching long actin~\cite{shingyoji98}
filaments to an ATPase molecule or microtubule~\cite{Kinoshita98}
filaments to a dynein molecule.  Both these assays rely on imaging the
probe to determine orientation, limiting the time resolution to video
rates on the order of 10 Hz.

Although it is inferred from the swimming speed of micro-organisms
that bacterial flagella can turn at a top rotational speed of
100,000~RPM~\cite{DeRosier98}, measurement techniques cannot keep up
with those rates.  Measurements are typically made by forcing the
flagella to turn a large moment of inertia which slows the rotation
rate to the region where it is accessible to video analysis.
Similarly, in studies of F$_1$-ATPase, the rotation rate is slowed to
an observable level by increasing the hydrodynamic drag that it must
oppose. However, the rotation is likely to be fundamentally at a
constant angle per ATP hydrolized, as shown by~\cite{noji97,yasuda98},
and slowing the rotation by increasing drag results in some elastic
storage of energy. Methods are needed that have the potential to
observe high-frequency rotations and have less hydrodynamic drag.

\begin{figure}
\epsfxsize=3.25in
\centerline{\epsfbox{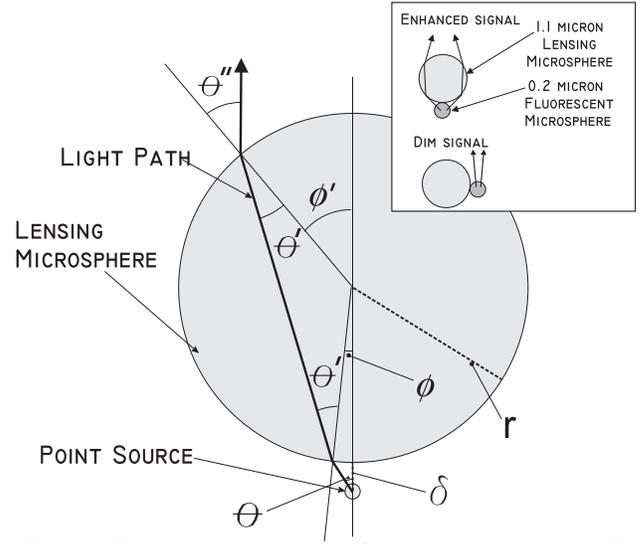}}
\caption{The optical ray tracing of the two microspheres.  This
  diagram defines the angles $\theta, \theta', \theta'', \phi, \phi'$
  and the distances $r$ and $\delta$.  The ray originally starts at an
  angle $\theta$ to the vertical and after passing through the lensing
  microsphere continues on at an angle $\phi' - \theta''$.  The index
  of refraction of the water is $n_1 = 1.3$ and for the polystyrene
  microsphere is $n_2=1.59$. Inset:  The fluorescence collected from an
  objective with finite NA is enhanced when the microsphere pair is
  aligned with the optical collection axis.}
\label{angles}
\end{figure}

This paper presents a method to measure high frequency rotational
motion.  An asymmetric fluorescent probe is made by attaching a small
fluorescently labeled polymer microsphere to a larger polymer
microsphere.  The larger sphere acts as a lens, substantially
enhancing the collection efficiency of the optical system.  The
experiment is outlined in the inset of Fig.~\ref{angles}.

We can show that the system increases the amount of collected light
with a geometric optics argument.  The angles of ray tracing are
outlined in Fig.~\ref{angles}; we would like to calculate the exit
angle $\phi' - \theta''$ as a function of the incident angle $\theta$.
The small fluorescent microsphere is approximated as a point particle
a distance $\delta$ from the lensing microsphere.  Using geometry, note that
\begin{equation}
\phi' = \pi - (\pi -2 \theta' + \phi) = 2 \theta' - \phi.
\end{equation}
Applying Snell's law at the top interface of the lensing microsphere
we find that
\begin{equation}
\theta'' = \sin^{-1}\left(\frac{n_2}{n_1} \sin \theta'\right),
\label{snellstop}
\end{equation}
where $n_2$ is the index of refraction of the lensing microsphere and
$n_1$ is the index of refraction of the surrounding medium (typically water).
Applying Snell's law at the bottom interface we obtain
\begin{equation}
\theta' = \sin^{-1} \left(\frac{n_1}{n_2} \sin (\theta + \phi)\right).
\label{snellsbottom}
\end{equation}
Then, by direct substitution of equation~(\ref{snellsbottom}) into
equation~(\ref{snellstop}), we find that
\begin{equation}
\theta'' = \phi + \theta.
\end{equation}

Using the law of sines, we can write
\begin{equation}
\frac{\sin (\pi - \theta - \phi)}{r+\delta} =\frac{\sin \theta}{r},
\end{equation}
and then explicitly find the angle $\phi$ as a function of $r, \theta$,
and $\delta$:
\begin{equation}
\phi(r,\theta,\delta) = \sin^{-1} \left( \frac{r + \delta}{r} \sin
  \theta \right) - \theta .
\end{equation}

Finally, we can write the exit angle $\phi' - \theta''$ in terms of
the original angle, $\theta$, the radii of the two spheres, $r,
\delta$, and the indices of refraction, $n_1$ and $n_2$.
\begin{equation}
\phi' - \theta'' = 2\sin^{-1}\left(\frac{n_1}{n_2} \sin \left(\theta +
    \phi(r,\theta,\delta) \right) \right)- \theta - 2 \phi(r,\theta,\delta). 
\end{equation}
For $\delta \ll r$, we note that $\phi \ll \theta $.  The exit angle is then given by
\begin{equation}
\phi' - \theta'' = 2\sin^{-1}\left(\frac{n_1}{n_2}\sin \theta\right) -
\theta.
\label{eight}
\end{equation}

Typical realizable values of $n_1$ and $n_2$ are for water, $n_1=1.3$
and polystyrene, $n_2=1.59$.  For small $\theta$ equation~(\ref{eight})
reduces to $\left( 2\frac{n_1}{n_2} - 1 \right) \theta$. This gives an
exit angle of $0.64\theta$ for a ray entering at an angle $\theta$.
Since the exit angle is always less than the original angle, we
conclude that the lensing microsphere focuses rays from the
fluorescent microsphere and enhances the optical signal.

The enhancement in the observed optical signal also depends on the
numerical aperture of the objective.  The numerical aperture (NA) is
defined as NA $=n \sin \theta_o$, where $\theta_o$ is the collection
angle.  For our objective (20$\times$, 0.4 NA) in air $\theta_o =
23.6\degree$.  Equation~(\ref{eight}) shows that the focusing
microsphere increases the angle of collection to $43.5\degree$.  This
corresponds to an effective NA of 0.69.  The epi-fluorescent intensity
is proportional to NA$^4$, so we expect an intensity enhancement of
$(0.69/0.4)^4\approx9$ times; this is of the order of what we
observed.  Objectives with high NA collect almost all of the emitted
light, and thus we would not expect to see any fluorescent enhancement
from a high NA objective.  Observations made with a 40\,$\times$,
1.30\,NA oil immersion objective indicate no noticeable intensity
enhancement.




To construct this system we used the bio\-tin/strep\-tavidin binding
system.  Biotin is known to bind streptavidin with a very high
affinity~\cite{wilchek90}.  We obtained streptavidin coated
microspheres from Interfacial Dynamics (polystyrene at a volume
concentration of 2.6\%, 1.1\,$\mu$m diameter) and biotin coated
yellow-green fluorescent microspheres (polystyrene 1\% by volume,
0.2\,$\mu$m diameter, excitation maximum at 505\,nm, emmission maximum
at 515\,nm) from Molecular Probes.  The smaller microspheres were
diluted by a factor of~1000 in distilled water.  A volume of
10\,$\mu$l of this solution was added to 10\,$\mu$l of the large
microspheres and allowed to incubate at room temperature for
approximately five minutes.
The solution was then diluted by a
factor of $10^4$ in either distilled water or a glycerol/water
mixture to give a final concentration of approximately one microsphere
pair per nanoliter at three different viscosities (1\,cP, 4\,cP, and 13\,cP).

\begin{figure}
\epsfxsize=3.25in
\centerline{\epsfbox{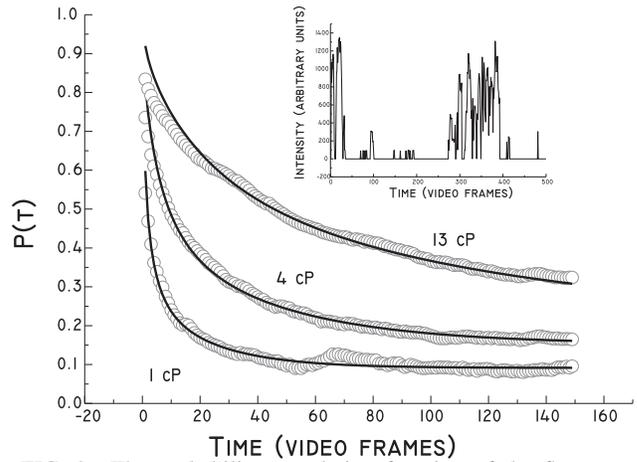}}
\caption{The probability correlation function of the fluorescent intensity
  signal measured in three different glycerol-water solutions of varying viscosities.
  These measurements are the average over approximately 30
  microspheres measured over approximately 1~minute.  The errors in
  the measurements are of order the diameter of the circles marking
  the data points.  The solid lines represent fits to the data with the
  following parameters $D_1= 1.30(\pm0.03)$\,s$^{-1}$,
  $\alpha_1=32.7(\pm0.2)^\circ$, $D_{4} = 0.70(\pm0.02)$\,s$^{-1}$,
  $\alpha_{4}=55.0(\pm0.5)^\circ$, $D_{13} =
  0.100(\pm0.003)$\,s$^{-1}$, $\alpha_{13} = 46.5(\pm1.1)^\circ$.  The
  inset shows typical data from a single microsphere ($\eta \approx
  4$\,cP) over 18 seconds.  The time scale is in video frames where a
  single video frame is $1/30$ of a second.}

\label{data}
\end{figure}

The solution was placed on a microscope slide with coverslip and the
edges were sealed with a clear nail polish to prevent any evaporation
or fluid flow.  These specimens were examined on a Nikon fluorescent
microscope with a 20\,$\times$, 0.4\,NA objective.  The lensing
microsphere is undergoing constant rotational diffusion, and when the
attached fluorescent microsphere is aligned with the optical
collection axis the observed intensity is substantially larger than in
other positions.

Images of these intensity fluctuations were observed using a CCD
camera and recorded on video tape.  The sequence of video images was
analyzed by digitizing them with a PC.  The microspheres' fluorescent
intensities were measured in each frame.  Data with better time
resolution could be obtained by using a photodiode to observe single
microspheres, but the video camera is sufficiently fast for observing
rotational diffusion and offered the advantage of being able to
observe many microsphere pairs in parallel.


From the intensity fluctuations we can compute the probability $P(t)$
that the intensity is above a threshold at time $t$, given that the
intensity was above this threshold at time $t=0$.  $P(t)$ can in turn
be calculated from first principles.  The rotational motion of the
lensing microsphere can be decomposed into orthogonal azimuthal and polar
directions.  Since only the azimuthal angle contributes to the
intensity enhancement, it suffices to consider a one dimensional
problem. Let $\alpha$ denote the azimuthal angle over which the signal
is enhanced, which then determines the initial conditions.  The
solution for the one dimensional diffusion equation with initial
condition $ |\theta| \le \alpha$ and infinite boundary conditions
is~\cite{crank70}
\begin{equation}
p_{inf}(\theta,t)= \frac{1}{2\alpha}\left(\mbox{erf}\left(\frac{\alpha/2 -
    \theta}{2\sqrt{D_r t}}\right)+ \mbox{erf}\left(\frac{\alpha/2 +
    \theta}{2\sqrt{D_r t}}\right) \right).
\end{equation}

This can be made periodic by summing $\theta$ over all multiples of $2\pi$.
The probability density function for the distribution in angles is then
\begin{equation}
p(\theta,t)=  \sum_{n=-\infty}^{n=\infty} p_{inf}(\theta + 2 n \pi,t)
\end{equation}
and the total probability distribution is 
\begin{equation}
P(t)= \int_{-\alpha/2}^{\alpha/2}p(\theta,t) d\theta.
\end{equation}

In practice this was approximated numerically using just the
largest terms of the series $(|n|\le3)$.
The $n=0$ term is computed using a Chebyshev approximation to
the error function and integrating it using a trapezoid
algorithm~\cite{numrecip}.  The higher order terms were computed using
the approximation that the initial condition was
$p(\theta,0)=\delta(\theta)$ (the solution is a gaussian
function) and the integration was carried out as the value of the
function times the width.

The data can therefore be fit to a two parameter function $P(t)$ that
is characterized by the angle $\alpha$ and the rotational
diffusion coefficient $D_r$.  The angle $\alpha$ is determined by
the optics of the system and the threshold applied to the data.  The
rotational diffusion coefficient $D_r$ is given by
\begin{equation}
D_r = \frac{k_B T}{8\pi \eta a^3},
\end{equation}
where $\eta$ is the viscosity and $a$ is the sphere's radius~\cite{vandeven}.
The measured diffusion coefficients are in agreement with the
predicted values (Fig.~\ref{data}).


We have outlined a mechanism by which a microlensing effect is used to
measure rotations of individual microspheres.  The microlensing can be
explained with geometric optics; a complete theory will need to take
into account near field optical effects.  The microlensing was
observed experimentally and used to measure the rotational diffusion
constant of a sphere.  This mechanism may have applications in studies
of biomolecular rotations and fluid dynamics, especially in situations
with a high rotational rate.
For a rotation at a constant rate, the load can be adjusted
by changing both the viscosity of the fluid and size of the larger
microsphere.  Precision microspheres are commercially available with
diameters of less than 1\micron\ to greater than 100\micron\ giving a
range of loads that differ by $10^6$.

This work was supported by the Whitaker Foundation 
and by NSF CAREER grant PHY-9722417.

\end{document}